\documentclass[aps,prl,reprint,groupedaddress]{revtex4-1}

\usepackage{amsmath}
\usepackage{amsfonts}
\usepackage[pdftex]{graphicx}
\usepackage[pdftex]{color} 
\begin{document}

%\preprint{}

\title{Kolmogorovian turbulence in transitional pipe flows}

\author{Rory T.\ Cerbus}
\author{Chien-chia Liu} 
\author{Gustavo Gioia} 
\author{Pinaki Chakraborty}
\affiliation{Okinawa Institute of Science and Technology 
Graduate University, Onna-son, Okinawa, Japan 904-0495}



\date{\today}

\begin{abstract}
As everyone knows who has opened a kitchen faucet, pipe flow is laminar at low flow velocities and turbulent at high flow velocities. At intermediate velocities there is a transition wherein plugs of laminar flow alternate along the pipe with ``flashes" of a type of fluctuating, non-laminar flow which remains poorly known. We show experimentally that the fluid friction of flash flow is diagnostic of turbulence. We also show that the statistics of flash flow are in keeping with Kolmogorov's phenomenological theory of turbulence (so that, e.g., the energy spectra of both flash flow and turbulent flow satisfy small-scale universality). We conclude that transitional pipe flows are two-phase flows in which one phase is laminar and the other, carried by flashes, is turbulent
in the sense of Kolmogorov.
\end{abstract}

%\keywords{}

\maketitle

%\section{}
%\subsection{}
%\subsubsection{}

In 1883, Osborne Reynolds \cite{reynolds1883experimental}
carried out a series of experiments to test the theoretical argument, 
rooted in the concept of dynamical similarity, that the character of a pipe flow should be controlled by the Reynolds number 
$Re \equiv UD/\nu$, where $U$ is the mean velocity of the flow, 
$D$ is the diameter of the pipe, and $\nu$ is the kinematic 
viscosity of the fluid. 
At low $Re$, Reynolds observed a type of flow, known as laminar, in which
``the elements of the fluid follow one another along lines of motion 
which lead in the most direct manner to their destination'' \cite{reynolds1883experimental}. 
At high $Re$, Reynolds observed a type of flow, 
known as turbulent, in which 
the elements of fluid 
``eddy about in sinuous paths the most indirect possible'' 
\cite{reynolds1883experimental}.
At intermediate values of $Re$, 
Reynolds found a transitional regime in which the flow was 
spatially heterogeneous, with plugs of steady, presumably laminar flow alternating with 
flashes (Reynolds's term) of fluctuating, eddying flow.

Since Reynolds's time much has been learned about
the transitional regime. It is now
known that there are two types of
flash \cite{wygnanski1973transition,wygnanski1975transition,
eckhardt2007turbulence, mullin2011experimental,barkley2016theoretical}: 
``puffs'' and ``slugs,''
where the puffs appear first, at the onset of the transitional regime,
which is triggered by finite-amplitude perturbations 
\cite{mullin2011experimental}, 
consistent
with Reynolds's finding that
the transitional regime starts at
a value of $Re$ that is highly sensitive to experimental 
conditions \cite{reynolds1883experimental}.
Shaped like downstream-pointing arrowheads 
\cite{wygnanski1973transition,mullin2011experimental},
the puffs %($Re < 2250$) 
are $\approx 20\, D$ long and travel 
downstream at a speed \cite{mullin2011experimental} of $\approx 0.9\, U$.
A puff may decay and disappear, %become steady, 
or it may split into two or three puffs
separated by intervening laminar plugs \cite{avila2011onset}.
The tendency to decay competes with the tendency to
split, which gains importance as $Re$ increases,
becoming dominant \cite{avila2011onset} at $Re \approx 2040$.
With further increase in $Re$, the puffs are replaced
by slugs \cite{barkley2015rise} ($Re \gtrsim 2250$). 
With a head faster than $U$ %the mean flow 
and a tail that is slower, the 
slugs, unlike the puffs, spread as they travel downstream and are
therefore capable of crowding out the laminar plugs.
As $Re$ increases further, the flow turns turbulent.

Although puff flow and slug flow
have been probed experimentally
\cite{wygnanski1973transition,wygnanski1975transition,
mullin2011experimental}, they
have scarcely being characterized statistically,
and at present it would be difficult to ascertain what
the most salient differences or similarities might be between,
say, the structure of puff flow and that of turbulent flow.
With these considerations in mind,
we 
start by turning our attention to $f$, 
a unitless measure of pressure drop per unit length of pipe.  Known as 
``fluid friction,'' 
$f$ is sensitive %diagnostic of
to the internal structure of a flow, and has therefore been used, 
ever since the time of Reynolds, to %ascertain whether a flow is 
classify flows as
laminar, turbulent, or transitional.

By definition, $f \equiv \frac{D \Delta P/\Delta L}{\rho U^2/2}$,
where $\Delta P/\Delta L$ is the pressure drop 
per unit length of pipe and
$\rho$ is the density of the fluid.
For laminar flow, Reynolds \cite{reynolds1883experimental}
found that $f = f_{\rm{lam}}(Re) \equiv 64/Re$,
%\propto Re^{-1}$, 
in accord with a classical mathematical result, derivable from the 
equations of motion of a flow dominated by viscosity. Fluctuations 
have a marked effect on $f$, and for turbulent flow 
Reynolds \cite{reynolds1883experimental}
found that $f=f_{\rm{turb}}(Re) \equiv 0.3164 Re^{-1/4}$ %\propto Re^{-1/4}$
(see Supplementary Text in Supplemental Material (SM)). 
This empirical result, known as the Blasius law
\cite{schlichting}, 
provides an excellent fit to experimental data
on turbulent flow for $Re < 10^5$.
The Blasius law has not yet
been derived from 
the equations of motion, but it has recently been shown that 
the scaling $f_{\rm{turb}}(Re) \propto Re^{-1/4}$
can be 
derived from Kolmogorov's phenomenological theory 
of turbulence \cite{GB02, GIOIA06}.
Lastly, for the transitional regime, Reynolds found that the 
relation between $f$ and $Re$ ``was either indefinite or very complex''
\cite{reynolds1883experimental}. 
Our immediate aim is to take another look at this complexity.

We carry out measurements of $f$ 
in a 20-m long, smooth, cylindrical glass pipe of 
$D = 2.5$ cm $\pm 10$ $\mu$m. % (see Methods). 
The fluid is water.
Driven by gravity, the flow remains laminar up to the highest $Re$ tested
($Re \approx 8300$).
To make flashes, we perturb the flow using
an iris or a pair of syringe pumps.
We compute data points ($f, Re$) by measuring,
at discrete times $t$, 
the mean velocity $U$ and
the pressure drop $\Delta P$ 
over a lengthspan $\Delta L = 202 D$.
The time series $\Delta P (t)$ and $U(t)$ 
yield $f(t)$ and $Re (t)$, which we
average over a long time ($> 4000$ $D/U$)
to obtain a single data point $(f, Re$).
(See Methods in SM for further discussion on the
experimental setup.)

\begin{figure*}
  \centerline{\includegraphics[scale=0.67]{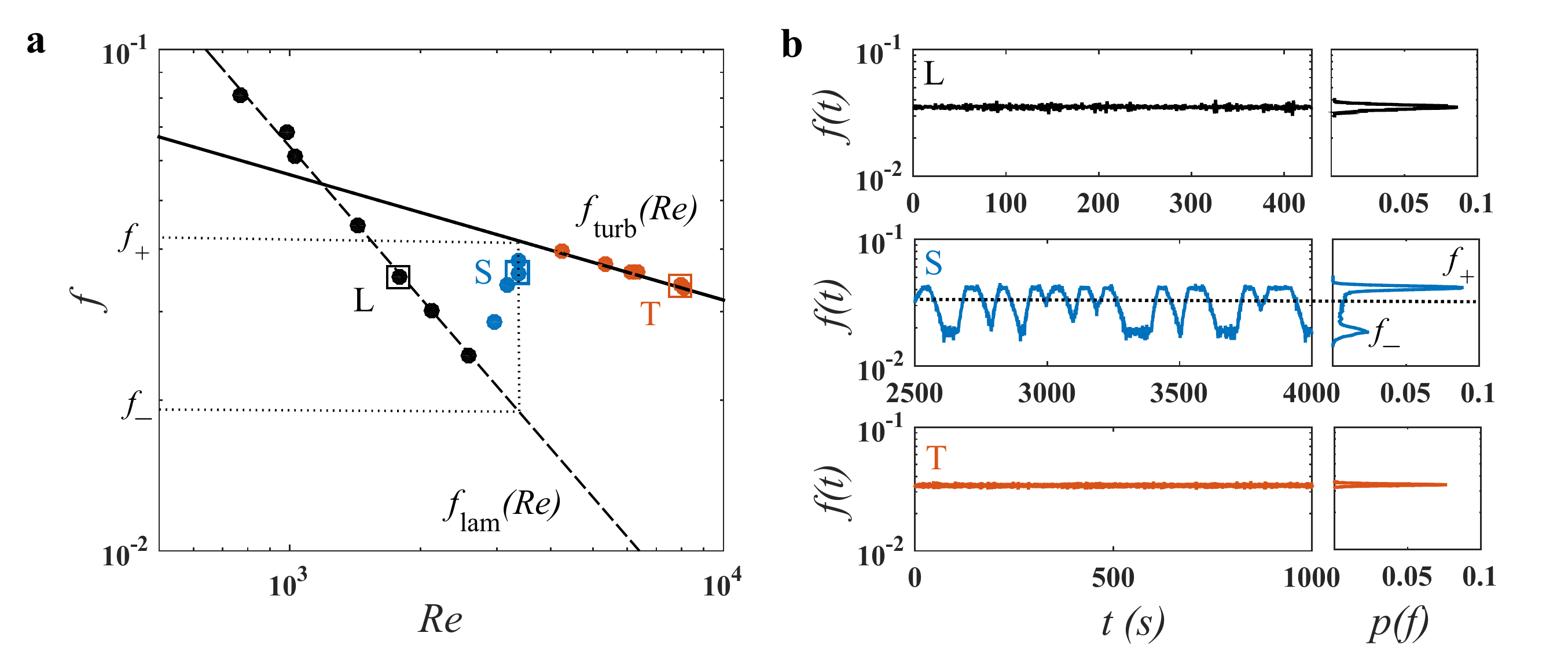}}
  \caption{\label{main1}
Experimental measurement of friction for transitional flows
with slugs. These flows consist of laminar plugs and slugs. 
%Here, all transitional flows are transitional flows with slugs. 
To make slugs, we perturb the flow using an iris. 
(a) Data points $(f, Re)$ corresponding to laminar flows
(shown in black), transitional flows (shown in blue), 
and turbulent flows (shown in red). 
The dashed line is the friction for laminar flow,
$f_{\rm{lam}}(Re)$;
the solid line is the friction for turbulent flow,
$f_{\rm{turb}}(Re)$ (the Blasius law).
(b) The time series $f(t)$ and the
attendant
probability distribution function $p(f)$ for the data points
marked L, S, and T in panel (a).
Note that for the transitional data point S, $f(t)$ swings between 
two distinct values, 
$f_{-}$ and $f_{+}$, where
%one corresponding to the
%laminar state and the other the turbulent state.
$f_{-} = f_{\rm{lam}}(Re)$ and $f_{+} = f_{\rm{turb}}(Re)$,
as indicated in panel (a).
Here, $f_{-}$ is the friction for laminar plugs, confirming that
laminar plugs are indeed laminar, and   
$f_{+}$ is the friction for slugs, suggesting that
slugs are turbulent. This conclusion holds for all the
transitional data points of panel (a).
}
\end{figure*}

In Fig.~\ref{main1}a, a log-log plot of
$f$ vs.\ $Re$, we see data points that
fall on $ f_{\rm{lam}}(Re)$ and correspond to
laminar flows;
data points that fall on $ f_{\rm{turb}}(Re)$ and
correspond to turbulent flows; and data points
that fall between $ f_{\rm{lam}}(Re)$ and $ f_{\rm{turb}}(Re)$
and correspond to transitional flows.
Here, all transitional flows are 
transitional flows with slugs. 
In Fig.~\ref{main1}b, we show the time series $f(t)$
and the attendant probability distribution function $p(f)$ for
a representative laminar data point (marked L), for a representative
transitional data point (marked S), %T$_{\rm{r}}$) 
and for a representative
turbulent data point (marked T).
For point L and for point T, $p(f)$ has a single peak, 
the value of which 
is the same as the long-time average of $f(t)$,
which we have denoted by $f$. 
By contrast, for the transitional 
data point S,
$p(f)$ has two peaks and
$f(t)$ swings \cite{durst2006forced} between the peak values, 
marked $f_{-}$ and $f_{+}$ in Fig.~\ref{main1}, 
spending little time at the long-time average of $f(t)$,
which %represents the friction for transitional flows and
includes contributions from both laminar plugs and
slugs. 
%(marked $f$ in Fig.~\ref{main1}b).
Here, $f_{-}$ is the friction for laminar plugs
%Here, the lower peak corresponds to laminar plugs
(that is, the unitless pressure drop
per unit length of laminar plug) and
%the higher peak corresponds to slugs
$f_{+}$ is the friction for slugs
(that is, the unitless pressure drop
per unit length of slug).
But there is more: it turns out that $f_{-} = f_{\rm{lam}}(Re)$
and $f_{+} = f_{\rm{turb}}(Re)$ (as indicated in Fig.~\ref{main1}a)---% 
and not just for data point S but
for all transitional data points in Fig.~\ref{main1}a.
This finding confirms
that laminar plugs are indeed laminar,
and, more important, it
suggests 
that
slugs are turbulent.

We now turn to transitional flows with puffs.
Unlike slugs, puffs are short
($\approx 20\, D$ long) as compared
to the lengthspan $\Delta L = 202\, D$,
and the technique we have used to measure 
%the pressure drop per unit length of slug 
$f$ for slugs
cannot be used for puffs.
%over which we measure the pressure drop.
To measure 
%the pressure drop per unit length of puff,
$f$ for puffs,
we create a train of puffs 
\cite{samanta2011experimental,hof2010eliminating}
such that at any given time
about 6-7 puffs fit within $\Delta L$. 
We then measure the time series $f(t)$, which
we average over a long time to obtain $f$. 
Now, this $f$ includes contributions
from both puffs and %intervening 
laminar plugs.
To disentangle
these contributions,
we use the procedure described in Methods (SM);
this procedure yields $f$ for puffs and
$f$ for laminar plugs.
Note that the same procedure can be applied to 
transitional flows with slugs,
%providing an opportunity to verify independently
%the results of Fig.~\ref{main1}.
as an alternative to the simpler procedure that we used
to obtain the results of Fig.~\ref{main1}b
(the results differ by $< 1.5\%$).

\begin{figure*}
  \centerline{\includegraphics[scale=0.68]{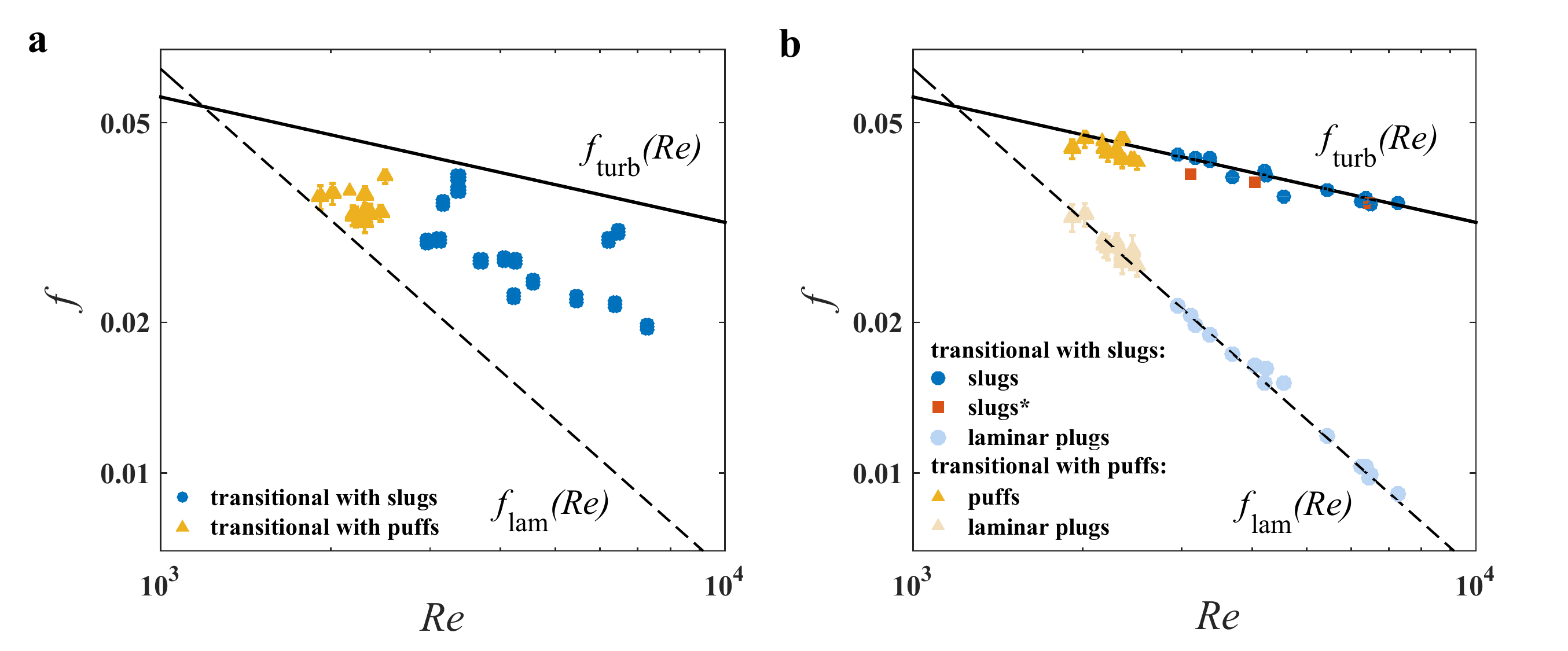}}
  \caption{\label{main3} 
Experimental measurement of friction for transitional flows.
% with slugs and transitional flows with puffs.
(a) Data points $(f, Re)$  corresponding to 
transitional flows with slugs (shown in blue)
and transitional flows with puffs (shown in yellow).
For each data point, $f$ includes 
a contribution from laminar plugs and a
contribution from flashes (either slugs or puffs, as the
case might be). By disentangling these contributions
 we obtain $f$ for laminar plugs ($f_{-}$)
and $f$ for flashes ($f_{+}$). In panel (b) we show the data point
  $(f_{-}, Re)$  and the data point $(f_{+}, Re)$
  for each and every data point in panel (a). 
In all cases, $f_{-}$ falls on $f_{\rm{lam}}(Re)$,
confirming that laminar plugs are indeed laminar,
and  $f_{+}$ falls on $f_{\rm{turb}}(Re)$, indicating that
flashes, irrespective of type, are turbulent.
All data points (panels a and b) have error bars. The vertical
error bars indicate errors in $f$ (see Supplementary 
Text in SM);
they are mostly smaller than the size of the data points. %at large $Re$. 
The errors in $Re$ are in all cases smaller than the size of the data points. 
%for all $Re$.
In panel (b), for the data points marked with $*$, we compute
$f$ for slugs using the same procedure that we use to compute $f$ for puffs.
}
\end{figure*}

In Fig.~\ref{main3}a, a log-log plot of $f$ vs.\ $Re$, 
we show all of the transitional data points that we have measured.
%including both slugs and puffs. 
For each of the transitional data points 
in Fig.~\ref{main3}a, 
we compute $f$ for flashes (either slugs or puffs,
as the case might be) and
$f$ for laminar plugs, and plot the
results in Fig.~\ref{main3}b. 
%For slugs, we use the
%procedure of Fig.~\ref{main1}b; for puffs, we use the
%procedure described in the previous paragraph.
For all transitional flows,
$f$ for laminar plugs equals
$f_{\rm{lam}}(Re)$ and
$f$ for flashes equals $f_{\rm{turb}}(Re)$,
irrespective of the type of flash.
Thus, by the conventions of fluid dynamics 
going back to the times of Reynolds, flash
flow is turbulent flow.

To adduce further experimental evidence that flash flow
is indeed turbulent flow, we turn to Kolmogorov's
phenomenological theory of turbulence \cite{KOLM41, FRIS95}, 
a theory that provides a thorough, empirically-tested description of the 
statistical structure of turbulent flow. 
Central to the phenomenological theory is the turbulent-energy spectrum $E(k)$,
which represents the way in which the turbulent kinetic energy is distributed among turbulent fluctuations of different wavenumbers $k$ in a flow.
Kolmogorov argued that, for $Re\rightarrow \infty$ 
and for $k$ in the ``universal range'' 
$k\gg D^{-1}$, $E(k)$ depends only on $k$, $\nu$, and 
$\varepsilon \propto U^3/D$, irrespective of the flow, where $\varepsilon$ is the turbulent power (that is, the rate at which turbulent kinetic energy is dissipated in the flow). In this case, Kolmogorov predicted that \cite{KOLM41, FRIS95}
\begin{equation}
E(k)\propto \frac{\nu^2}{\eta} F(k\eta),
\label{k1}
\end{equation}
which is known as small-scale universality,
and
\begin{equation} 
\eta \propto D Re^{-3/4},
\label{k2}
\end{equation}
where $F$ is a universal function and $\eta$ is 
the viscous lengthscale.
Further, for $k$ in the ``inertial range'' $D^{-1}\ll k\ll \eta^{-1}$,
 a subset of the universal range, 
$E(k)\propto \varepsilon^{2/3} k^{-5/3}$,
% \begin{equation}
%E(k)\propto \varepsilon^{2/3} k^{-5/3},
%\end{equation}
which is known as the ``5/3 law.'' 
%Kolmogorov assumed $Re\gg 1$. 
Eq.~\eqref{k1}, 
Eq.~\eqref{k2}, and the 5/3 law
are asymptotic results associated with the limit $Re \rightarrow \infty$, and it is not possible to predict 
mathematically for what finite value of $Re$ they might be expected to hold within a preset tolerance. 
In practice, it is only feasible to verify the 5/3 law over a broad inertial range, which necessitates a small 
ratio $\eta/D$, which in turn necessitates a particularly large value of $Re$ (due to the small exponent of $Re$ in 
Eq.~\eqref{k2}). Indeed, in pipe flows the 5/3 law becomes clearly apparent \cite{rosenberg2013turbulence} only for 
$Re>80,000$, well 
above the values of $Re$ at which flashes have been observed. By contrast, 
Eq.~\eqref{k1} and Eq.~\eqref{k2} 
can hold in principle
  at the values of $Re$ of our experiments (also see
\cite{schumacher2007asymptotic, schumacher2014small}). Note, however, that whereas the 5/3 law can be tested by 
carrying out a single experiment at a very high value of Re, a test of Eq.~\eqref{k1} and Eq.~\eqref{k2} 
requires that many experiments be carried out over a broad range of values of $Re$.

\begin{figure*}
%\vspace{-0.5cm}
  \centerline{\includegraphics[scale=0.7]{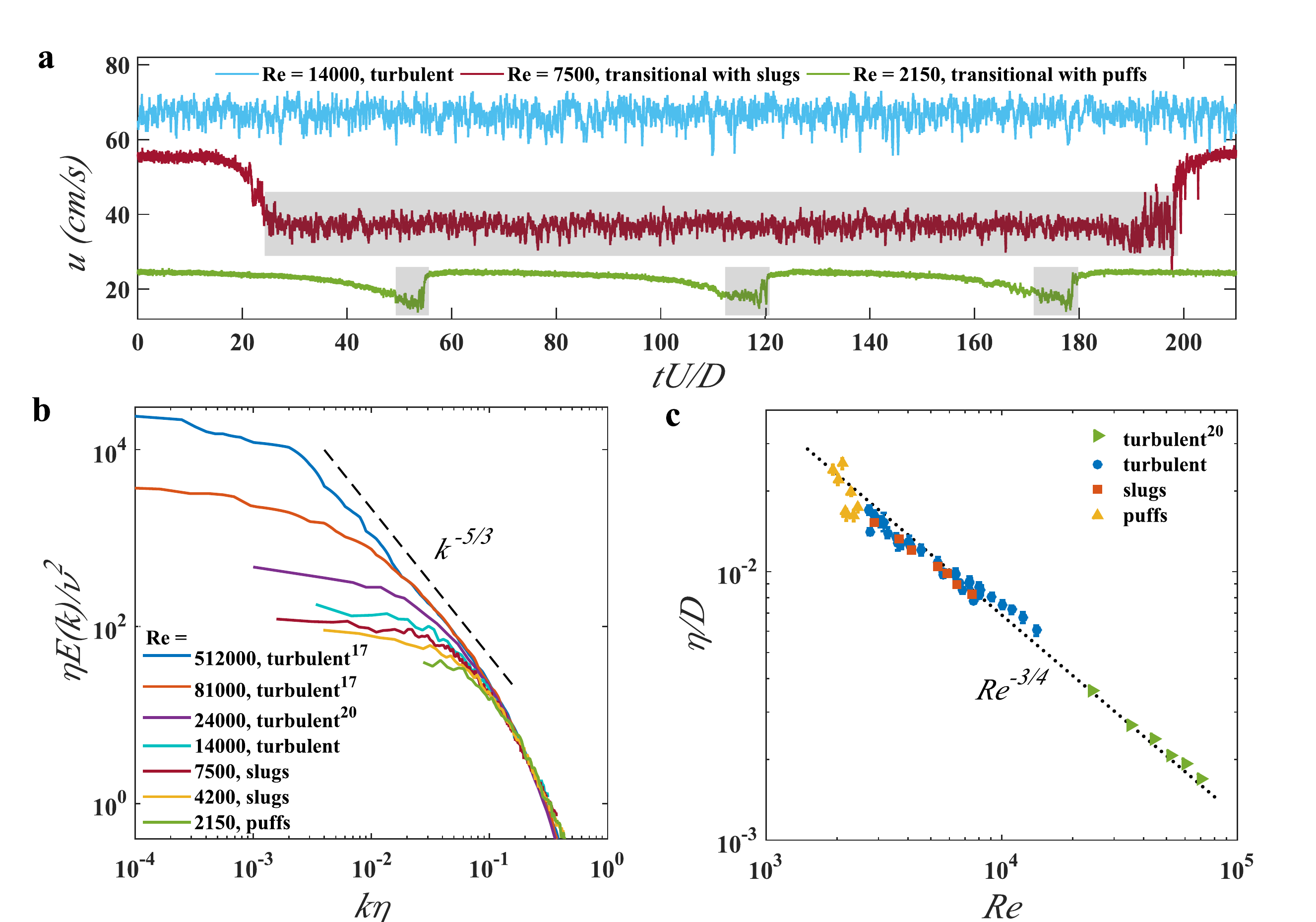}}
%\vspace{-0.5cm}
  \caption{\label{eta}
Tests of Kolmogorov's phenomenological theory of turbulence,
Eqs.~\eqref{k1} and \eqref{k2}, for flash flows.
(a) Time series of the axial velocity $u$ at the centerline
of the pipe for three representative flows. 
(For the transitional flow with puffs, we plot 
$1.5\, u(t)$ for the sake of clarity.)
%For the transitional flows 
The segments of $u(t)$ that correspond to slugs
and puffs have been shaded in grey. We use
those segments to compute the spectra $E(k)$ and 
the viscous lengthscale $\eta$ for slug flow
and for puff flow.
 (b) Plots of the rescaled spectra
$\eta E(k)/\nu^2$ vs.\ $k \eta$
for a few representative flows, including
turbulent flows and flash flows.
The rescaled spectra are in good keeping
with small-scale universality (Eq.~\eqref{k1}),
irrespective of the type of flow.
(c) Data points $(\eta/D, Re)$ for all our
experiments and for a few high-$Re$ Princeton
superpipe experiments \cite{bailey2009measurement}, 
showing agreement with the Kolmogovorian scaling of Eq.~\eqref{k2}.
%(where we increased the highest $Re$ to $\approx 14,000$
%using a pump)
}
\end{figure*}

To compute $E(k)$ we start by measuring
%Using an LDV we measure 
a time series of the axial velocity  
at the centerline of the pipe, $u(t)$, 
using laser Doppler velocimetry (LDV). 
 In Fig.~\ref{eta}a we plot $u(t)$ for
three representative flows: 
a turbulent flow, 
a transitional flow with slugs, and
a transitional flow with puffs; for the
transitional flows, the segments of $u(t)$
corresponding to flashes have been shaded in grey. 
%turbulent,
%transitional with slugs (there the segments shaded in grey
%correspond to slug flow), and transitional with puffs
%(there the segments shaded in grey
%correspond to puff flow).
Using time series such as those of
Fig.~\ref{eta}a, we compute $E(k)$ and $\eta$
for several turbulent flows, 
slug flows, and puff flows
(see Methods in SM). 
In Fig.~\ref{eta}b, we show a few representative spectra $E(k)$,
along with a few high-$Re$ spectra from 
the Princeton superpipe experiment
\cite{bailey2009measurement, rosenberg2013turbulence}.
The spectra of Fig.~\ref{eta}b have been rescaled in
order to verify small-scale universality (Eq.~\eqref{k1}).
At high $k \eta$, the rescaled spectra $\eta E(k)/\nu^2$ vs.\ $k \eta$
converge onto the universal function $F(k \eta)$ of Eq.~\eqref{k1} for
turbulent flows as well as for flash flows, in keeping with
small-scale universality. Further, the rescaled spectra peel
off from $F(k \eta)$ at a value of $k \eta$ that lessens
monotonically as $Re$ increases, irrespective of the type of flow.

In Fig.~\ref{eta}c, we test Eq.~\eqref{k2} using data points
from all our experiments along with a few high-$Re$
data points from the Princeton superpipe experiment
\cite{bailey2009measurement}.
Irrespective of the type of flow, the data points,
which span about two decades in $Re$,
are in keeping with Eq.~\eqref{k2}.
From Figs.~\ref{eta}b and \ref{eta}c, we conclude that
the spectra and the viscous lengthscales
of flash flows, like those of conventional
turbulent flows, 
are governed by the phenomenological
theory of turbulence, with the implication
that flash flows,
aside from their being 
restricted to relatively low $Re$,
are statistically indistinguishable
from conventional turbulent flows.

In a number of experimental and computational studies 
\cite{sano2016universal,lemoult2016directed,shih2016ecological}
published in 2016, compelling evidence has been adduced in support of a 30 year-old conjecture 
by Yves Pomeau \cite{pomeau1986front} to the effect that the subcritical transition in pipe 
flow and other shear flows belongs to the directed-percolation universality class of non-equilibrium 
phase transitions. %\cite{barkley2016theoretical}. 
Yet, in a comment on those %experimental and computational 
studies, titled ``The Long and Winding Road,'' Pomeau 
\cite{pomeau2016long}
cautioned that 
``the arrowhead patterns observed in early experiments 
[on boundary layers] are sufficiently regular to denote a 
bifurcation to a turbulence-free 
state.''
That is to say, if flashes were non-turbulent, they could hardly be the agents of a transition 
to turbulence, and the experimental and computational evidence of directed percolation would be 
severed from the turbulent regime. Our findings indicate that, at least for pipe flow, flashes display 
fluid-frictional behavior diagnostic of turbulence and a statistical structure indistinguishable 
from that of conventional turbulent flow %sensu
in the sense of 
Kolmogorov. Thus, we conclude that flash flow is but 
turbulent flow, and flashes endow the transitional regime with 
the requisite link to turbulence.
Our findings suggest 
that new insights into the transition to turbulence
may be gained 
by approaching the transition from above,
from higher to lower $Re$,
complementing the 
usual approach from below.
The long road keeps winding.

\begin{acknowledgments}
We thank Prof.\ Jun Sakakibara (Meiji University) and 
Mr.\ Makino (Ni-gata company)
for help with the experimental setup.
This work was supported by
the Okinawa Institute of Science and 
Technology Graduate University.
\end{acknowledgments}

\end{document}